%% file: foellmi.tex
\documentclass[multphys,vecphys]{svmult}

\usepackage{makeidx}         
\usepackage{graphicx}        
\usepackage{multicol}        
\usepackage[bottom]{footmisc}

\makeindex             

\begin{document}

\title*{Bisectors as Distance Estimators for Microquasars?}
\author{C. Foellmi\inst{1,2}}
\institute{European Southern Observatory, 3107 Alonso de Cordova, Vitacura, Casilla 19.,  Santiago, Chile
\texttt{cfoellmi@eso.org}
\and
Laboratoire d'Astrophysique, Observatoire de Grenoble, 414 rue de la Piscine, 38400 Saint-Martin d'H\`eres, France
\texttt{cfoellmi@obs.ujf-grenoble.fr}}
%
%
\maketitle

\section{Scientific Context}
\label{sec:1}

Microquasars \index{microquasars} are small-scaled version of their extragalactic parents, the quasars\index{quasars} with which they share the same physics \cite{Mirabel-2004}. They have a central black-hole\index{black-hole} (here of stellar mass), an accretion disk and (sometime persistent) jets\index{jets}. These jets are sometimes relativistic, and can reach apparent superluminal motion. However, the superluminal "property" depends on the distance\index{distance} between the object and the observer.

Recently, the distance of the second known superluminal microquasar in our Galaxy, namely GRO J1655-40\index{GRO J1655-40} \cite{Hjellming-Rupen-1995}, has been challenged by \cite{Foellmi-etal-2006b-astroph}. At a new distance smaller than 1.7~kpc (instead of 3.2~kpc), the jets are not superluminal anymore. It is thus of central importance to obtain reliable values of the distance in order to achieve an consistent understanding of these relativistic objects. At an inferred distance of 1.0~kpc, GRO J1655-40 competes with the microquasar called 1A 0620-00\index{1A 0620-00} to be the closest black-hole to the sun.

\section{New Exploratory Method to Obtain Distance Estimate}
\label{sec:2}

We propose here a new and exploratory method\index{method} to obtain an {\it estimate} of the distance of microquasars. A distance estimate, say within 10-20\%, is already a great progress for microquasars, where distances are sometimes uncertain by a factor of 2. This new method is made of two ingredients.

First, it has been shown recently by \cite{Gray-2005} that the bluemost point of the spectroscopic bisector\index{bisector} of a star is linearly related to its absolute magnitude\index{absolute magnitude}. It is an empirical relation working (at least) for late-type stars. Unfortunately, to obtain a good quality {\it single-line} bisector, it is necessary to achieve a Signal-to-Noise ratio of at least 300 (often even higher) in the considered line, and at the same time a high spectral resolution (say, above 40 000). For faint objects in the optical such as GRO J1655-40 (V $\sim$ 18), a bisector to be made with UVES\index{UVES} (R=45 000) would require more than two continuous weeks of 24-hours observations...

Second, it has been shown recently by \cite{Dall-etal-2006} using the HARPS\index{HARPS} instrument installed in the La Silla Observatory (Chile)), that the bisector of the cross-correlation function\index{cross-correlation function} (CCF) can be used as much the same way as single-line bisectors. This is rather natural, since well-chosen atmospheric lines should behave mostly the same way.

The combination of the two ingredients opens the door to the observation of faint targets. As a matter of fact, it is much easier to obtain a good quality bisector of the CCF, since the information of numerous lines can be combined together. The Signal-to-Noise ratio required now depends also on the number of lines used to compute the CCF.

\section{Known Caveats and Prospects}
\label{sec:3}

The idea behind this method is to obtain a good-quality spectrum of the secondary star of microquasars in quiescence, compute their CCF and measure the position of the bluemost point of it. This point being then compared to the linear relation as described in \cite{Dall-etal-2006}. Of course, there are obvious caveats in this method that we list here.
\begin{itemize}
\item The correlation between (CCF) bisector and absolute magnitude is not fully understood yet. 
\item The correlation is known to work only for late-type stars, in addition to have been tested on a small sample only. 
\item Microquasars have often Roche-filling secondary stars that could influence in an unclear manner the bisector and the absolute magnitude.
\item The spectrum of microquasar can be polluted by black-hole accretion disk\index{accretion disk}
\end{itemize}

Despite these caveats, we are currently investigating this relation using the newly commissionned near-infrared echelle spectrograph\index{echelle spectrograph} at the VLT\index{VLT}: CRIRES\index{CRIRES}. With such instrument, it is possible to achieve the necessary high resolution, and obtain a good quality spectrum of microquasars in the near-infrared, where microquasars are much brighter than in the optical (for GRO J1655-40, K$\sim$12.5).

\input{referenc}

\printindex
\end{document}

%% file: referenc.tex
%
%

%
%